\documentclass[useAMS,usenatbib]{mn2e}

\usepackage{epsfig}
\usepackage{multirow}
\usepackage{array}
\usepackage{color}
\usepackage{amsmath}

\usepackage{url}
\usepackage{hyperref}

\title[Revised H$_2$ cooling function]{Temperature and density dependent cooling function for H$_2$ with updated H$_2$/H collisional rates}
\author[Carla Maria Coppola et al.]{Carla Maria Coppola$^{1,2*}$\thanks{E-mail: carla.coppola@uniba.it}, Fran\c{c}ois Lique$^{3}$, Francesca Mazzia$^{4}$, Fabrizio Esposito$^{5}$
\newauthor and Mher V. Kazandjian$^{6*}$ \\
$^1$Universit\`{a} degli Studi di Bari, Dipartimento di Chimica, Via Orabona 4, 
I-70126, Bari, Italy\\
$^2$INAF-Osservatorio Astrofisico di Arcetri, Largo E.~Fermi 5, I-50125, Firenze
, Italy\\
$^3$LOMC - UMR 6294, CNRS-Universit\'e du Havre, 25 rue Philippe Lebon, BP 1123,
 76 063 Le Havre cedex, France\\
$^4$Universit\`{a} degli Studi di Bari, Dipartimento di Informatica, Via Orabona 4, 
I-70126, Bari, Italy\\
$^5$Consiglio Nazionale delle Ricerche, Via Amendola 118, I-70126, Bari, Italy\\
$^6$Sterrewacht Leiden, Leiden University, PO Box 9513, 2300 RA, Leiden, The Netherlands\\
$^*$These authors equally contributed to this work.
}

\begin{document}

\pagerange{\pageref{firstpage}--\pageref{lastpage}} \pubyear{2014}

\maketitle

\label{firstpage}

% \shorttitle{Revised H$_2$ cooling function}
% \shortauthors{Coppola et al.}
% 
% 
% \begin{document}
% 
% \title{Revised H$_2$ cooling function}
% 

% \author{Carla Maria Coppola}
% \affiliation{Universit\`a degli Studi di Bari, Dipartimento di Chimica, Via Orabona 4, I-70126 Bari, Italy}
% \affiliation{INAF-Osservatorio Astrofisico di Arcetri, Largo E.~Fermi 5, I-50125 Firenze, Italy}
% \author{Mher Vatche Kazandjian}
% \affiliation{Sterrewacht Leiden, Leiden University, PO Box 9513, 2300 RA, Leiden, The Netherlands}
% \author{Fran\c{c}ois Lique}
% \affiliation{LOMC - UMR 6294, CNRS-Universit\'e du Havre, 25 rue Philippe Lebon, BP 1123, 76 063 Le Havre cedex, France}

\begin{abstract}
  The energy transfer among the components in a gas determines its fate.
  Especially at low temperatures, inelastic collisions drive the cooling and the heating 
  mechanisms. In the early Universe as well as in
  zero- or low- metallicity environments the major contribution comes from
  the collisions among atomic and molecular hydrogen, also in
  its deuterated version. The present work shows some updated calculations of the
  H$_2$ cooling function based on novel collisional data which explicitely take into account the 
  reactive
  pathway at low temperatures. Deviations from previous calculations are discussed and a
  multivariate data analysis is performed to provide a fit depending on both the gas
  temperature and the density of the gas.
\end{abstract}


\section{Introduction}
The thermal evolution of an interstellar cloud with primordial composition is deeply driven by the 
collisional and radiative processes occuring in the medium. In order to be properly described, it is
essential not only to correctly model the dynamics but also to provide a complete description of the 
time 
variation of the gas temperature. To address this need, it is firstly required to identify the 
possible
cooling mechanisms in the intergalactic medium (IGM); secondly, the models seek for functions that can 
be
easily adopted in the models to reproduce each of the heating/cooling channels. In the context of IGM
simulations and low metallicity interstellar medium (ISM) several works have been produced adopting 
this 
workflow\citep[e.g.,][]{black1981,shapiro_hydrogen1987,cen_hydrodynamic1992,glover2007,glover2008,glover2009,glover2015}; the 
same holds in the case of early Universe chemistry \citep{galli2013}, where the system is mainly consisting 
on hydrogen and helium.
Mechanical ways of energy exchange are found in shocks and turbulence \cite[e.g.,][]{johnson_cooling2006,kazandjian2016,kazandjian2016b,vasiliev2017a}. The most 
relevant coolants at low 
temperatures are metals and molecules, both in gas phase and as dust in the intergalacting and in the 
interstellar medium. According to the radiation field in which the gas and the dust is embedded, 
heating mechanisms should be also taken into account in the thermal balance (e.g., \cite{vasiliev2017b}).
%\citep[e.g., ][]{vasiliev2017b}.
On cosmological scales, the presence of even small 
traces of metals can determine a 
different fate in terms of masses of the first stars because of the line cooling \citep[e.g.,][]
{bromm2003,yoshida2007}; gravitational instabilities are expected in the case of increasing 
metallicity content \citep{tanaka_gravitational2014}, together with thermal instabilities 
\citep{omukai_thermal_inst_2015}. Indeed, the original gravitationally collapsing cloud can fragment 
or keep 
accreting according to the thermal conditions present in the gas \citep{omukai2000,latif2016}. At the 
limit, 
the presence of coolant agents (both atomic and molecular) and a radiation field which has an 
intensity 
below the critical value can allow for the formation of supermassive black holes via direct accretion 
\citep[e.g.,][]
{haiman_h1996,sugimura2016}. In the absence of heavier elements like carbon and oxygen (and the 
consequent 
possibility to form molecules like CO) the only way to cool the gas down to temperatures below few 
hundreds 
Kelvin is through rovibrational transitions of H$_2$ and its isotopic variants, the latter allowing to 
cool 
the gas even at temperatures below 100~K. For this reason, the description of the energy transfer 
associated 
with the H$_2$ line cooling deserves to be rigorously described, also by adopting the most complete 
data for each of the possible collisional partners. Specifically, in this paper, the case of the 
inelastic 
collisions of H$_2$ with atomic hydrogen is studied, explicitely introducing the reactive channel.
The work is organized as follows: in Section~\ref{methods} the basic concepts of the collisional 
pathways in 
the H$_3$ system are provided, together with the quantum mechanical methods adopted in the calculation 
of 
the cross-sections and the equation for the calculation of the cooling function. The kinetic model adopted and the chemical pathways included are also described;
in Section~ \ref{multivariate}, the results are reported. Moreover, a regression analysis of the data obtained for the cooling function is performed 
to derive a multivariate dependence of the cooling function on the gas temperature and density.

\section{Methods and equations.}
\label{methods}

In order to describe the time evolution of the level population for a $N$-levels system, the rate
equation for each level should be provided. In the high density regime, the levels are distributed 
according to 
Maxwell-Boltzmann (e.g., \cite{coppola2011_radiative}).  This hypothesis is in general not valid and 
the level population
can be found by explicitely solving the rate equations which derive from the chemical processes 
introduced in the kinetics.
In order to find the level population, two general approaches are usually adopted: on one hand, the 
time evolution of the levels population $\mathbf{x}$ can be
found by solving the resulting system of ODEs
\cite{coppola2011a,coppola2011b}; otherwise, it can be assumed that the level population is in steady-state
\cite{martin1996,tine1998,coppola2012}. The calculations reported in this paper have been performed 
using
the second assumption, that corresponds to imposing that the time derivative of the rate equations is equal to 0:
\begin{equation}
\frac{dx_i}{dt} = 0, ~~ \forall i.
\label{steady}
\end{equation}
At each temperature and density, the whole problem of finding the steady-state population
can be translated into a linear system formalism. Let $\mathbf{M}$ be the matrix
that, when multiplied by the column vector of the population densities $\mathbf{n}$, results in the right 
hand side of the system of ordinary differential equations; the rate equations can be written then as \citep{tine1998}:
\begin{equation}
\frac{d\mathbf{n}}{dt} = \mathbf{M} \cdot \mathbf{n}\\
\label{odes}
\end{equation}
Then, the level population derives from the chemical channels that most effectively redistribute them among the rovibrational manifold; the choice of such channels 
depends on the system of interest. The case reported in this paper corresponds to the typical freeze-out abundances that can be found in the early Universe 
chemistry. The 
H$_2$ collisional partners included are H, He and H$^+$; moreover, the formation and destruction pathways for the molecular hydrogen are the associative detachment 
of H and H$^-$ and the dissociative attachment of H$_2$, respectively. Radiative transition are inserted and the data by \cite{simbotin1998} have been adopted. The 
relative abundances of He and H$^+$ respect to H are 10$^{-1}$ and 2$\times10^{-4}$. The reaction rates 
included for the formation and destruction channels of H$_2$ have been taken from \cite{coppola2011a} and \cite{coppola2007}. The collisional data with He are 
fully available for the whole H$_2$ rovibrational manifold \citep[F. Esposito, private communication and][]{esposito2017}. The collisional data for the system H
$_2$-H$^+$ have been included using the fits provided by \cite{gerlich1990}; transitions up to ($v=0, j=8$) are therein reported. Finally, the reactive H$_2$-H 
collisions have been included; details on the adopted data are described in the following. According to the reactive data provided for H$_2$-H, the number of 
levels included in the present calculations is 55, that corresponds to ($v=3, j=18$).

\subsection{Updates collisional data: H$_2$--H reaction rates.}
The modeling of accurate cooling functions relies on the calculations of H$_2$-H collisional rate coefficients. Such calculations are challenging since two
 processes are in competition during collisions between H$_2$ and H:\\
-the inelastic process:
$$
\mathrm{H}_2(v,j) + \mathrm{H}' \to \mathrm{H}_2\mathrm{(v',j') + H'}
\label{process1}
$$\\
-the exchange process:
$$
\mathrm{H}_2(v,j) + \mathrm{H}' \to  \mathrm{H'H(v',j') + H}
\label{process2}
$$
where $v$ and $j$ designates the vibrational and rotational level of H$_2$, respectively. This is explained by the reactive nature of the H$_3$ system. The 
obtention of accurate collisional data requires to consider simultaneously both processes during the calculations. In particular, the ortho--para-H$_2$ conversion 
can only occur through the hydrogen exchange channel.

Ro-vibrational relaxation of H$_2$ by H have been extensively studied using quasi-classical trajectory calculations \cite[][and references therein]{Mandy:93}. 
Unfortunately, the inability of quasi-classical trajectory treatments to conserve the vibrational zero-point energy renders this method unreliable near reaction 
thresholds. Alternatively, Flower and co-workers \citep{Wrathmall07,Wrathmall07b} computed ro-vibrational rate coefficients for temperatures ranging from 100 to 
6000~K using a quantum close coupling approach neglecting the reactive channels arguing that reactivity is negligible for temperatures up to 6000~K. They did not 
consider the ortho--para-H$_2$ conversion process and neglect the vibrational relaxation that occur through the exchange process that is expected to be important 
even at low temperatures. 

Hence, there were still a lack of highly accurate collisional data until we recently presented quantum mechanical calculations of cross sections for the 
collisional excitation of H$_2$ by H including the reactive channels \citep{Lique:12:2,Lique:14,lique:15} using the state-of-the-art PES of \cite{Mielke:02}. 
We refer the reader to these papers for full details on the scattering calculations. In summary, calculations were performed using the almost exact close coupling 
approach. New collisional data were obtained for the ro-vibrational relaxation of highly excited H$_2$ (with internal excitation up to $\simeq$ 22000~K) for 
temperatures ranging from 100 to 5000~K. The theoretical results were checked against available experimental data and a good agreement was found for both ro-
vibrational relaxation \citep{lique:15} and ortho--para-H$_2$ conversion process \citep{Lique:12:2}.

The new results significantly differ from previous data widely used in astrophysical models \citep{Wrathmall07,Wrathmall07b}. Important deviations are observed at 
low temperatures for ro-vibrational transitions whereas, for pure rotational transitions, the mean deviations occur at high temperatures. These differences are 
principally due to the inclusion of the reactive channels in the scattering calculations.

\subsection{Cooling function}
\label{cooling}
The calculation of the cooling function requires information about the level population, computed at each gas temperature, and on the energy gaps and Einstein 
coefficients between the levels included in the model. In particular, for a generic molecule $mol$, the cooling function in defined as:
\begin{equation}
n_{mol} \Lambda = \sum_j \sum_{i<j} n_j A_{ji}\beta_{ji}(E_j - E_i)
\label{cooling_function_expression}
\end{equation}
with $\Lambda$ in erg s$^{-1}$, n$_{mol}$ the density of the molecule, $n_j$ the density for the $j^{th}$ 
level and $\beta_{ji}$ is the escape probability of the emitted photon. The assumption under which the 
calculations here presented are performed is that the medium is 
optically thin; in this case, all the emitted photons can escape from the medium without being re-absorbed 
and in Eq.~\ref{cooling_function_expression} the term $\beta_{ji}$ is equal to 1. The calculation of the 
cooling function has been performed using the code \texttt{FRIGUS} 
developed 
by \cite{kazandjian2019}. No assumptions have been adopted on the level population between ortho- and
para-H$_2$, that have been explicitely calculated according to the steady-state approximation and detailed 
balance between radiative and collisional rates. Stimulated process are included in 
\texttt{FRIGUS} where a Planckian radiation field has been implemented by default to mimic the cosmic 
microwave background at a certain redshift $z$; however, the calculations reported in this work have been 
performed for a radiation temperature equal to zero. The 
Einstein coefficients have been calculated by \cite{simbotin1998}; the energy levels used have been taken 
from the \href{https://www.physast.uga.edu/ugamop/mainh2.html}{UGA - Molecular Opacity Project Database}
and the collisional reaction rates adopted are the ones computed by \cite{lique:15}.
% The derivation of these equations can be found in the
% \href{https://github.com/mherkazandjian/frigus/blob/master/doc/rate_equations_derivation.ipynb} available in 
% the \texttt{FRIGUS}'s \texttt{GitHub} repository.

\section{Results}
\label{multivariate}

\subsection{Cooling function}
\begin{figure}
\includegraphics[width = 0.5\textwidth]{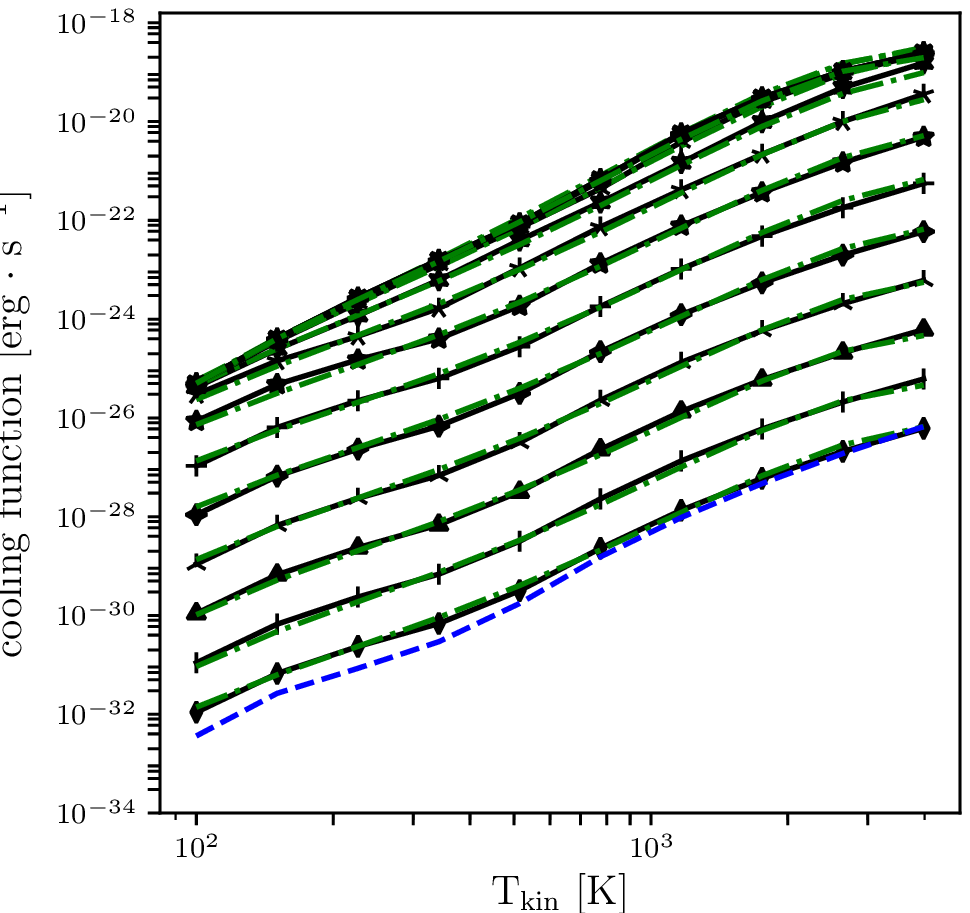}
\caption{Cooling functions at different densities adopting the new data \citep{Lique:12:2,Lique:14,lique:15} 
and the software \texttt{FRIGUS} (black continuous curves). From bottom to top, the 
corresponding 
densities are:
10$^{2}$ m$^{-3}$, 10$^{3}$ m$^{-3}$, 10$^{4}$ m$^{-3}$, 10$^{5}$ m$^{-3}$, 10$^{6}$ m$^{-3}$, 10$^{7}$ m$^{-3}$, 10$^{8}$ m$^{-3}$, 10$^{9}$ m$^{-3}$, 10$^{10}$ m$^{-3}$, 10$^{11}$ m
$^{-3}$, 10$^{12}$ m$^{-3}$, 10$^{13}$ m$^{-3}$, 10$^{14}$ m$^{-3}$
% . 
% For the same densities, the cooling 
% functions computed adopting the data by Wrathmall \& Flower~2008 are also reported (dash-dotted green lines)
As 
a comparison, the fit provided by Glover \& Abel~2008 is also shown (blue dashed curve). The dash-dotted curves correspond to 
the fit provided in the present paper; see text for details.}
\label{cf}
\end{figure}

In Fig.~\ref{cf} the cooling function obtained adopting the updated collisional reaction rates by 
\cite{lique:15} is reported as a function of the kinetic temperature for several values of the gas density 
(black continuous curves, from the low to the high density limit, corresponding to the lower and upper curves, respectively).
% A 
% comparison is provided with the cooling functions obtained using the software \texttt{FRIGUS} and adopting 
% the data by \cite{Wrathmall07} (dash-dotted green curves in Fig.~\ref{cf}). 
% The introduction of reactive channels in the calculation of the cross-sections for the inelastic processes 
% allows for interesting cooling effects that depend on the temperature range and on the density. In 
% particular, for each density, it is possible to find a temperature at which the cooling function calculated 
% with the new collisional data becomes smaller than the cooling function derived adopting the data by 
% \cite{Wrathmall07}. This effect is due to a systematic overestimate of the level populations at low 
% temperatures when using the data by \cite{Wrathmall07}, caused by having neglected the ortho-to-para 
% conversion. For the same reason, as the temperature exceeds these crossing values, at each density, 
% the cooling with the new collisional 
% data becomes larger, because hydrogen exchange is now allowed.
The dependence from the density follows what shown by \cite{lipovka2005} for HD: the curves tend to converge 
for high values of density to 
the LTE cooling function, i.e. to the cooling function that can be obtained by assuming that the level 
population is described by a Maxwell-Boltzmann distribution for the energy levels. This case corresponds to 
a gas in which the fractional abundances of internal rovibrational levels is thermalized, while in the more 
general case this situation is not necessarly satisfied. 

\begin{figure}
\includegraphics[width = 0.5\textwidth]{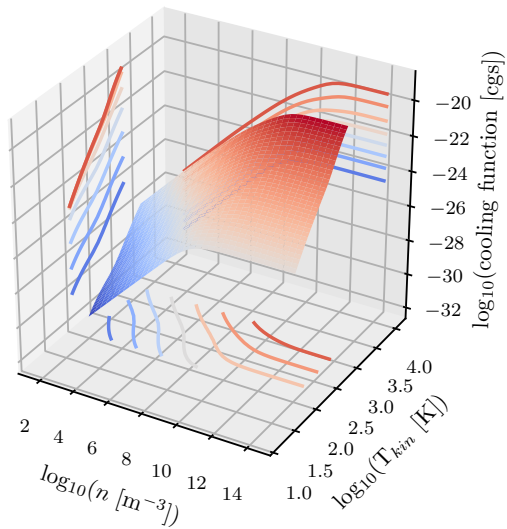}
\caption{3D rendering of the cooling function at different densities and kinetic temperatures using \texttt{FRIGUS} (Kazandjian \& Coppola 2019). SI units are used.}
\label{3d}
\end{figure}
In the same figure, the most recent fit for H$_2$/H 
cooling by \cite{glover2008} is also reported, which is based on the data by \cite{Wrathmall07}.
At low temperatures T$\sim 100~K$, 
deviations up to an order of magnitude can be appreciated.
Together with the different collisional data implemented, 
such a difference may be explained assuming that their fit 
has been performed assuming an ortho-to-para ratio 3:1.
%; in the following, we will 
% identify the ortho-to-para ratio with the acronym ``opr''. 
In the absence of a full state-to-state description of the kinetics of rovibrational levels, assuming an ortho-to-para  
is the only way to proceed in the calculations. Moreover, collisional processes with other atomic or 
molecular partners may allow to reach the statistical ortho-to-para in a faster way, allowing to confidently use the 3:1 
value in the simulations. For example, it is well 
known that collision with protons are very effective in the ortho-to-para conversion of H$_2$, as reported by 
\cite{gerlich1990} and recently confirmed by \cite{grozdanov2014}. However, as 
also acknowledged in the introduction of the work by \cite{glover2008}, deviations from the 3:1 ortho-to-
para ratio are expected at low temperatures, as explicitely captured by performing the calculations using 
\texttt{FRIGUS}). The ideal procedure would then prescribe to solve the kinetics of rovibrational level of H$_2$ in each particular case and to compute the actual level population and cooling {\it on-the-fly}.

In Fig.~\ref{3d} a 3D-rendering of the cooling functions surface is reported with a colour code to 
distinguish more easily the values of the independent variables and the cooling functions themselves.

\subsection{Fit}
\label{section_fit}
In order to allow for a faster usage of the calculated cooling function at several kinetic temperatures and densities, we provide the users with a fitted expressions for the cooling functions; the same expression used by \cite{lipovka2005} in the case of HD is used:
\begin{equation}
log10(\Lambda) = \sum_{l, m = 0}^4 D_{lm} (log_{10}{T})^l (log_{10}{n})^m
\end{equation}
where the numerical values of the logarithms are taken expressing the temperature in Kelvin and the 
density in cgs units; the resulting cooling function $\Lambda$ has units erg $\times$ s$^{-1}$. The 
parameters are provided in Table~\ref{fit_parameters}. In Fig.~\ref{cf} the comparison between the 
computed data (black full) and the fit (green dashed curves) is also reported, showing very small 
errors (the largest being $\sim$10$^{-3}$).

\begin{table*}
\centering
  \begin{tabular}{@{}cccccc@{}}
  \hline
  \hline
        &   m = 0          & m = 1            & m = 2          & m = 3          & m = 4     \\
  \hline
  l = 0 & -1.07761178e+02 & +8.50704285e+00  & -3.08850636e-01 & -3.97357274e-01 & +3.65589231e-02\\
  l = 1 & +1.17901741e+02  & -1.09489742e+01 &  +2.51708582e-01 & +5.47452438e-01 & -4.89055533e-02\\
  l = 2 & -6.61721218e+01 &+5.95232129e+00   & -1.10639748e-01  &-2.92423859e-01  &+2.57375696e-02 \\
  l = 3 &+1.67152116e+01   &-1.43513388e+00  & +2.84953800e-02 &+7.02040444e-02  &-6.18478233e-03 \\
  l = 4 & -1.55475014e+00 &+1.29639563e-01   &-3.13943372e-03   & -6.36261310e-03 &+5.65797161e-04 \\  
 \hline
 \hline
\label{fits}
\end{tabular}
\caption{Parameters for the cooling function fit as reported in Sec.~\ref{section_fit}; $\Lambda$ has units erg $\times$ s$^{-1}$. These parameters are valid in the density range [10$^2$-10$^{14}$]~m$^{-3}$ and temperature range [100-4000]~K.}
\label{fit_parameters}
\end{table*}

%\begin{figure}
% \includegraphics[width = 0.5\textwidth]{images/9.eps}
% \includegraphics[width = 0.5\textwidth]{images/7.eps}
% \caption{Comparison between computed cooling functions (continuous lines) and the corresponding values obtained using the fit provided in this paper (dashed line); density range as in Fig.~\ref{cf}.}
% \label{cf_fit}
% \end{figure}

% \begin{figure}
% \includegraphics[width=8.5cm]{}
% \caption{cooling_function_H2 at several gas densities. The figure reports the calculations obtained using the updated database by \cite{lique:15}} 
% \label{cooling_function_H2}
% \end{figure}

\section{Conclusions}
\label{conclusions}
Thanks to available updated cross-sections for the inelastic processes H$_2(v,j)$+H $\rightarrow$ H$_2(v',j')$+H \citep{lique:15} and the state-to-state software 
\texttt{FRIGUS} \citep{kazandjian2019}, a new cooling function depending both on gas temperature and density has been evaluated, which include the rovibrational 
levels up to ($v=3, j=18$). The inclusion of reactive channels increased the cooling effect at high temperatures, together with a larger set of included 
rovibrational levels in the energy exchange evaluation
% These effects are due to the 
% ortho-to-para ratio obtained by solving the rate equations and adopting the data by \cite{lique:15}, that in the case of low temperature and low densities can be 
% as high as 0.3.

For facilitating the usage of these results, a multivariate analysis is performed and the resulting fit is reported together with the parameters in Sec.~
\ref{section_fit}.  The computation has been performed for radiation temperature equal to zero, to compare the results with previous calculations; however, if 
needed, the software \texttt{FRIGUS} can be used to recover the cooling function at different radiation temperature (i.e. different epochs in the case of 
simulations describing early Universe chemistry or early star formation).

\section{acknowledgments*}
C.M.C greatly acknowledges Regione Puglia for the project ``Intervento cofinanziato dal Fondo di 
Sviluppo e Coesione 2007-2013 – APQ Ricerca Regione Puglia - Programma regionale a sostegno della 
specializzazione intelligente e della sostenibilit\`a sociale ed ambientale - FutureInResearch''. 
C.M.C. also acknowledges Daniele Galli who strongly encouraged the calculations reported in this work.

%\facility{facility ID}
%\facilities{facility ID, facility ID, facility ID} 
%\software{Numpy}

%\bibliographystyle{elsarticle-num}
\bibliographystyle{mn2e}
\bibliography{biblio,h3,mine}

\begin{thebibliography}{41}
\expandafter\ifx\csname natexlab\endcsname\relax\def\natexlab#1{#1}\fi

\bibitem[{{Black}(1981)}]{black1981}
{Black} J.~H., 1981, \mnras, 197, 553

\bibitem[{{Bromm} \& {Loeb}(2003)}]{bromm2003}
{Bromm} V., {Loeb} A., 2003, Nature, 425, 812

\bibitem[{{Capitelli} {et~al}\mbox{.}(2007){Capitelli}, {Coppola}, {Diomede},
  \& {Longo}}]{coppola2007}
{Capitelli} M., {Coppola} C.~M., {Diomede} P., {Longo} S., 2007, \aap, 470, 811

\bibitem[{{Celiberto} {et~al}\mbox{.}(2017){Celiberto}, {Capitelli}, {Colonna},
  {D'Ammando}, {Esposito}, {Janev}, {Laporta}, {Laricchiuta}, {Pietanza},
  {Rutigliano}, \& {Wadehra}}]{esposito2017}
{Celiberto} R. {et~al.}, 2017, Atoms, 5, 18

\bibitem[{Cen(1992)}]{cen_hydrodynamic1992}
Cen R., 1992, The Astrophysical Journal Supplement Series, 78, 341

\bibitem[{Coppola {et~al}\mbox{.}(2012)Coppola, D'Introno, Galli, Tennyson, \&
  Longo}]{coppola2012}
Coppola C., D'Introno R., Galli D., Tennyson J., Longo S., 2012, Astrophysical
  Journal, Supplement Series, 199

\bibitem[{Coppola {et~al}\mbox{.}(2011{\natexlab{a}})Coppola, Lodi, \&
  Tennyson}]{coppola2011_radiative}
Coppola C., Lodi L., Tennyson J., 2011{\natexlab{a}}, Monthly Notices of the
  Royal Astronomical Society, 415, 487

\bibitem[{Coppola {et~al}\mbox{.}(2011{\natexlab{b}})Coppola, Longo, Capitelli,
  Palla, \& Galli}]{coppola2011a}
Coppola C., Longo S., Capitelli M., Palla F., Galli D., 2011{\natexlab{b}},
  Astrophysical Journal, Supplement Series, 193

\bibitem[{{Galli} \& {Palla}(2013)}]{galli2013}
{Galli} D., {Palla} F., 2013, \araa, 51, 163

\bibitem[{{Gerlich}(1990)}]{gerlich1990}
{Gerlich} D., 1990, \jcp, 92, 2377

\bibitem[{{Glover}(2015)}]{glover2015}
{Glover} S.~C.~O., 2015, \mnras, 451, 2082

\bibitem[{{Glover} \& {Abel}(2008)}]{glover2008}
{Glover} S.~C.~O., {Abel} T., 2008, \mnras, 388, 1627

\bibitem[{Glover \& Jappsen(2007)}]{glover2007}
Glover S. C.~O., Jappsen A.-K., 2007, The Astrophysical Journal, 666, 1

\bibitem[{{Glover} \& {Savin}(2009)}]{glover2009}
{Glover} S.~C.~O., {Savin} D.~W., 2009, \mnras, 393, 911

\bibitem[{{Grozdanov}(2014)}]{grozdanov2014}
{Grozdanov} T.~P., 2014, in Journal of Physics Conference Series, Vol. 488,
  Journal of Physics Conference Series, p. 012030

\bibitem[{Haiman {et~al}\mbox{.}(1996)Haiman, Rees, \& Loeb}]{haiman_h1996}
Haiman Z., Rees M.~J., Loeb A., 1996, The Astrophysical Journal, 467, 522

\bibitem[{{Inoue} \& {Omukai}(2015)}]{omukai_thermal_inst_2015}
{Inoue} T., {Omukai} K., 2015, \apj, 805, 73

\bibitem[{Johnson \& Bromm(2006)}]{johnson_cooling2006}
Johnson J.~L., Bromm V., 2006, Monthly Notices of the Royal Astronomical
  Society, 366, 247

\bibitem[{Kazandjian {et~al}\mbox{.}(2016a)Kazandjian, Pelupessy, Meijerink,
  Israel, Coppola, Rosenberg, \& Spaans}]{kazandjian2016}
Kazandjian M., Pelupessy I., Meijerink R., Israel F., Coppola C., Rosenberg M.,
  Spaans M., 2016a, Astronomy and Astrophysics, 595

\bibitem[{Kazandjian \& Coppola(2019)}]{kazandjian2019}
Kazandjian M.~V., Coppola C.~M., 2019, in preparation

\bibitem[{{Kazandjian} {et~al}\mbox{.}(2016b){Kazandjian}, {Pelupessy},
  {Meijerink}, {Israel}, \& {Spaans}}]{kazandjian2016b}
{Kazandjian} M.~V., {Pelupessy} I., {Meijerink} R., {Israel} F.~P., {Spaans}
  M., 2016b, \aap, 595, A125

\bibitem[{{Latif} {et~al}\mbox{.}(2016){Latif}, {Omukai}, {Habouzit},
  {Schleicher}, \& {Volonteri}}]{latif2016}
{Latif} M.~A., {Omukai} K., {Habouzit} M., {Schleicher} D.~R.~G., {Volonteri}
  M., 2016, \apj, 823, 40

\bibitem[{{Lipovka} {et~al}\mbox{.}(2005){Lipovka}, {N{\'u}{\~n}ez-L{\'o}pez},
  \& {Avila-Reese}}]{lipovka2005}
{Lipovka} A., {N{\'u}{\~n}ez-L{\'o}pez} R., {Avila-Reese} V., 2005, Monthly
  Notices of the Royal Astronomical Society, 361, 850

\bibitem[{{Lique}(2015)}]{lique:15}
{Lique} F., 2015, \mnras, 453, 810

\bibitem[{{Lique} {et~al}\mbox{.}(2012){Lique}, {Honvault}, \&
  {Faure}}]{Lique:12:2}
{Lique} F., {Honvault} P., {Faure} A., 2012, \jcp, 137, 154303

\bibitem[{Lique {et~al}\mbox{.}(2014)Lique, Honvault, \& Faure}]{Lique:14}
Lique F., Honvault P., Faure A., 2014, Int. Rev. in Phys. Chem., 33, 125

\bibitem[{Longo {et~al}\mbox{.}(2011)Longo, Coppola, Galli, Palla, \&
  Capitelli}]{coppola2011b}
Longo S., Coppola C., Galli D., Palla F., Capitelli M., 2011, Rendiconti
  Lincei, 22, 119

\bibitem[{{Mandy} \& {Martin}(1993)}]{Mandy:93}
{Mandy} M.~E., {Martin} P.~G., 1993, \apjs, 86, 199

\bibitem[{{Martin} {et~al}\mbox{.}(1996){Martin}, {Schwarz}, \&
  {Mandy}}]{martin1996}
{Martin} P.~G., {Schwarz} D.~H., {Mandy} M.~E., 1996, \apj, 461, 265

\bibitem[{{Mielke} {et~al}\mbox{.}(2002){Mielke}, {Garrett}, \&
  {Peterson}}]{Mielke:02}
{Mielke} S.~L., {Garrett} B.~C., {Peterson} K.~A., 2002, \jcp, 116, 4142

\bibitem[{{Omukai}(2000)}]{omukai2000}
{Omukai} K., 2000, \apj, 534, 809

\bibitem[{Shapiro \& Kang(1987)}]{shapiro_hydrogen1987}
Shapiro P.~R., Kang H., 1987, The Astrophysical Journal, 318, 32

\bibitem[{Shchekinov \& Vasiliev(2017)}]{vasiliev2017b}
Shchekinov Y.~A., Vasiliev E.~O., 2017, Astrophysics, 60, 449

\bibitem[{Sugimura {et~al}\mbox{.}(2016)Sugimura, Coppola, Omukai, Galli, \&
  Palla}]{sugimura2016}
Sugimura K., Coppola C., Omukai K., Galli D., Palla F., 2016, Monthly Notices
  of the Royal Astronomical Society, 456, 270

\bibitem[{Tanaka \& Omukai(2014)}]{tanaka_gravitational2014}
Tanaka K. E.~I., Omukai K., 2014, {arXiv}:1401.2993 [astro-ph]

\bibitem[{{Tin{\'e}} {et~al}\mbox{.}(1998){Tin{\'e}}, {Lepp}, \&
  {Dalgarno}}]{tine1998}
{Tin{\'e}} S., {Lepp} S., {Dalgarno} A., 1998, Mem. S.A.It, 69, 345

\bibitem[{Vasiliev \& Shchekinov(2017)}]{vasiliev2017a}
Vasiliev E.~O., Shchekinov Y.~A., 2017, Astronomy Reports, 61, 342

\bibitem[{Wolniewicz {et~al}\mbox{.}(1998)Wolniewicz, Simbotin, \&
  Dalgarno}]{simbotin1998}
Wolniewicz L., Simbotin I., Dalgarno A., 1998, The Astrophysical Journal
  Supplement Series, 115, 293

\bibitem[{{Wrathmall} \& {Flower}(2007)}]{Wrathmall07}
{Wrathmall} S.~A., {Flower} D.~R., 2007, J. Phys. B, 40, 3221

\bibitem[{{Wrathmall} {et~al}\mbox{.}(2007){Wrathmall}, {Gusdorf}, \&
  {Flower}}]{Wrathmall07b}
{Wrathmall} S.~A., {Gusdorf} A., {Flower} D.~R., 2007, \mnras, 382, 133

\bibitem[{{Yoshida} {et~al}\mbox{.}(2006){Yoshida}, {Omukai}, {Hernquist}, \&
  {Abel}}]{yoshida2007}
{Yoshida} N., {Omukai} K., {Hernquist} L., {Abel} T., 2006, \apj, 652, 6

\end{thebibliography}

\end{document}